\documentclass[runningheads]{llncs}

\usepackage[T1]{fontenc}
\usepackage{graphicx}
\usepackage{booktabs}
\usepackage{multirow}
\usepackage{amsmath}
\usepackage{url}
\usepackage[hidelinks]{hyperref}
\usepackage{xcolor}

\usepackage[T1]{fontenc}
\usepackage{graphicx}
\newcommand{\nPreTotal}{199}
\newcommand{\nPostTotal}{136}
\newcommand{\nPreM}{99}
\newcommand{\nPostM}{65}
\newcommand{\nPreF}{100}
\newcommand{\nPostF}{71}
\newcommand{\nPreStraight}{5}
\newcommand{\nPostStraight}{9}

\newcommand{\rqiAIKnownM}{98}
\newcommand{\rqiAIKnownF}{99}
\newcommand{\rqiAIKnowMM}{2.46}
\newcommand{\rqiAIKnowMF}{2.40}
\newcommand{\rqiAIKnowB}{-0.04}
\newcommand{\rqiAIKnowP}{.793}
\newcommand{\rqiAIKnowD}{-.04}
\newcommand{\rqiUndTrainnM}{98}
\newcommand{\rqiUndTrainnF}{98}
\newcommand{\rqiUndTrainMM}{3.00}
\newcommand{\rqiUndTrainMF}{2.71}
\newcommand{\rqiUndTrainB}{-0.25}
\newcommand{\rqiUndTrainP}{\textbf{.046}}
\newcommand{\rqiUndTrainD}{-.29}
\newcommand{\rqiUndBiasnM}{99}
\newcommand{\rqiUndBiasnF}{98}
\newcommand{\rqiUndBiasMM}{3.25}
\newcommand{\rqiUndBiasMF}{3.08}
\newcommand{\rqiUndBiasB}{-0.12}
\newcommand{\rqiUndBiasP}{.359}
\newcommand{\rqiUndBiasD}{-.13}

\newcommand{\rqiConsentnM}{98}
\newcommand{\rqiConsentnF}{97}
\newcommand{\rqiConsentMM}{3.42}
\newcommand{\rqiConsentMF}{3.35}
\newcommand{\rqiConsentB}{-0.03}
\newcommand{\rqiConsentP}{.799}
\newcommand{\rqiConsentD}{-.04}
\newcommand{\rqiAIConfnM}{77}
\newcommand{\rqiAIConfnF}{86}
\newcommand{\rqiAIConfMM}{2.69}
\newcommand{\rqiAIConfMF}{2.48}
\newcommand{\rqiAIConfB}{-0.23}
\newcommand{\rqiAIConfP}{.096}
\newcommand{\rqiAIConfD}{-.27}

\newcommand{\rqiCareerEngnM}{81}
\newcommand{\rqiCareerEngnF}{80}
\newcommand{\rqiCareerEngMM}{2.86}
\newcommand{\rqiCareerEngMF}{1.94}
\newcommand{\rqiCareerEngB}{-0.91}
\newcommand{\rqiCareerEngP}{\textbf{{<}.001}}
\newcommand{\rqiCareerEngD}{-.81}
\newcommand{\rqiCareerCSnM}{80}
\newcommand{\rqiCareerCSnF}{79}
\newcommand{\rqiCareerCSMM}{2.31}
\newcommand{\rqiCareerCSMF}{1.61}
\newcommand{\rqiCareerCSB}{-0.67}
\newcommand{\rqiCareerCSP}{\textbf{{<}.001}}
\newcommand{\rqiCareerCSD}{-.68}
\newcommand{\rqiCareerAInM}{79}
\newcommand{\rqiCareerAInF}{82}
\newcommand{\rqiCareerAIMM}{2.06}
\newcommand{\rqiCareerAIMF}{1.55}
\newcommand{\rqiCareerAIB}{-0.49}
\newcommand{\rqiCareerAIP}{\textbf{{<}.001}}
\newcommand{\rqiCareerAID}{-.58}

\newcommand{\MotSchoolPctM}{49.5}
\newcommand{\MotSchoolPctF}{65.0}
\newcommand{\MotSchoolOR}{1.81}
\newcommand{\MotSchoolP}{.045}
\newcommand{\MotFunPctM}{48.5}
\newcommand{\MotFunPctF}{40.0}
\newcommand{\MotFunOR}{0.71}
\newcommand{\MotFunP}{.233}
\newcommand{\MotProblemPctM}{27.3}
\newcommand{\MotProblemPctF}{30.0}
\newcommand{\MotProblemOR}{1.21}
\newcommand{\MotProblemP}{.546}
\newcommand{\MotAdvicePctM}{14.1}
\newcommand{\MotAdvicePctF}{27.0}
\newcommand{\MotAdviceOR}{2.35}
\newcommand{\MotAdviceP}{.021}

\newcommand{\DfSeenPctM}{71.7}
\newcommand{\DfSeenPctF}{74.0}
\newcommand{\DfSeenOR}{1.08}
\newcommand{\DfSeenP}{.805}
\newcommand{\DfSharedPctM}{23.2}
\newcommand{\DfSharedPctF}{10.0}
\newcommand{\DfSharedOR}{0.34}
\newcommand{\DfSharedP}{.010}
\newcommand{\DfCreatedPctM}{10.1}
\newcommand{\DfCreatedPctF}{4.0}
\newcommand{\DfCreatedOR}{0.32}
\newcommand{\DfCreatedP}{.067}
\newcommand{\DfOfMePctM}{12.1}
\newcommand{\DfOfMePctF}{6.0}
\newcommand{\DfOfMeOR}{0.42}
\newcommand{\DfOfMeP}{.101}

\newcommand{\CopilotD}{.57}

\newcommand{\rqiiMAIKnowDeltaM}{+0.61}
\newcommand{\rqiiMAIKnowP}{\textbf{{<}.001}}
\newcommand{\rqiiMAIKnowD}{.63}
\newcommand{\rqiiMUndTrainDeltaM}{+0.29}
\newcommand{\rqiiMUndTrainP}{.051}
\newcommand{\rqiiMUndTrainD}{.32}
\newcommand{\rqiiMUndBiasDeltaM}{+0.15}
\newcommand{\rqiiMUndBiasP}{.321}
\newcommand{\rqiiMUndBiasD}{.16}

\newcommand{\rqiiMConsentDeltaM}{+0.01}
\newcommand{\rqiiMConsentP}{.962}
\newcommand{\rqiiMConsentD}{.01}
\newcommand{\rqiiMAIConfDeltaM}{+0.52}
\newcommand{\rqiiMAIConfP}{\textbf{{<}.001}}
\newcommand{\rqiiMAIConfD}{.58}
\newcommand{\rqiiMCareerAIDeltaM}{-0.08}
\newcommand{\rqiiMCareerAIP}{.645}
\newcommand{\rqiiMCareerAID}{.08}
\newcommand{\rqiiMCareerCSDeltaM}{+0.10}
\newcommand{\rqiiMCareerCSP}{.613}
\newcommand{\rqiiMCareerCSD}{.09}
\newcommand{\rqiiMCareerEngDeltaM}{+0.37}
\newcommand{\rqiiMCareerEngP}{.073}
\newcommand{\rqiiMCareerEngD}{.31}

\newcommand{\rqiiIDMNetflixChange}{+27.2}
\newcommand{\rqiiIDMNetflixP}{{<}.001}
\newcommand{\rqiiIDMSpotifyChange}{+25.3}
\newcommand{\rqiiIDMSpotifyP}{.001}

\newcommand{\rqiiFAIKnowDeltaM}{+0.51}
\newcommand{\rqiiFAIKnowP}{\textbf{{<}.001}}
\newcommand{\rqiiFAIKnowD}{.57}
\newcommand{\rqiiFUndTrainDeltaM}{+0.67}
\newcommand{\rqiiFUndTrainP}{\textbf{{<}.001}}
\newcommand{\rqiiFUndTrainD}{.75}
\newcommand{\rqiiFUndBiasDeltaM}{+0.42}
\newcommand{\rqiiFUndBiasP}{\textbf{.002}}
\newcommand{\rqiiFUndBiasD}{.49}

\newcommand{\rqiiFConsentDeltaM}{+0.28}
\newcommand{\rqiiFConsentP}{\textbf{.049}}
\newcommand{\rqiiFConsentD}{.31}
\newcommand{\rqiiFAIConfDeltaM}{+0.60}
\newcommand{\rqiiFAIConfP}{\textbf{{<}.001}}
\newcommand{\rqiiFAIConfD}{.64}
\newcommand{\rqiiFCareerAIDeltaM}{+0.45}
\newcommand{\rqiiFCareerAIP}{\textbf{.002}}
\newcommand{\rqiiFCareerAID}{.52}
\newcommand{\rqiiFCareerCSDeltaM}{+0.44}
\newcommand{\rqiiFCareerCSP}{\textbf{.005}}
\newcommand{\rqiiFCareerCSD}{.47}
\newcommand{\rqiiFCareerEngDeltaM}{+0.24}
\newcommand{\rqiiFCareerEngP}{.182}
\newcommand{\rqiiFCareerEngD}{.22}

\newcommand{\rqiiIDFNetflixChange}{+28.2}
\newcommand{\rqiiIDFNetflixP}{{<}.001}
\newcommand{\rqiiIDFSpotifyChange}{+17.8}
\newcommand{\rqiiIDFSpotifyP}{.011}
\newcommand{\rqiiIDFTikTokChange}{+12.8}
\newcommand{\rqiiIDFTikTokP}{.016}

\begin{document}
\title{Gender Differences in AI Literacy Workshop Outcomes and Deepfake Engagement}
\titlerunning{Gender Differences in AI Literacy Workshop Outcomes}

\author{Jake Renzella\inst{1}\orcidID{0000-0002-9587-1196} \and
Christian Bergh\inst{1}\orcidID{0009-0000-2424-7356} \and
Natasha Banks \inst{2}\orcidID{0009-0008-8264-1990} \and
Alexandra Vassar \inst{1}\orcidID{0000-0001-8856-2566}}
\authorrunning{J. Renzella et al.}
\institute{University of New South Wales, Sydney \\ \href{mailto:jake.renzella@unsw.edu.au}
%{jake.renzella@unsw.edu.au} 
\and Day of AI Australia
}
%\author{Anon Author 1\inst{1}\orcidID{1111-2222-3333-4444} \and
%Anon Author 2\inst{1}\orcidID{1111-2222-3333-4444} \and
%Anon Author 3\inst{1,2}\orcidID{1111-2222-3333-4444} \and
%Anon Author 4\inst{2}\orcidID{1111-2222-3333-4444}
%}
%
%\authorrunning{Anon Author 1 et al.}
%\institute{Anon Institution 1 \and
%Anon Institution 2}
%
\maketitle              % typeset the header of the contribution

\begin{abstract}
As Artificial Intelligence (AI) literacy initiatives expand in K--12 settings, understanding how gender shapes student baseline perceptions, tool-use, and responsiveness to interventions is essential for equitable curriculum design.
This study examines gender differences in AI literacy, safety awareness, and STEM career aspirations among Australian secondary students (Years~7, 8, and~10; $N_{\text{pre}}=\nPreTotal$, $n_{\text{post}}=\nPostTotal$) from two co-educational government schools who participated in a one-day AI literacy workshop.
Using statistical regression methods controlling for year level and school, we found that pre-workshop, male students reported significantly higher STEM career interest across all three domains (AI, computer science, and engineering), while female students were significantly more likely to use AI for schoolwork and to seek advice from AI tools.
Gender-differentiated patterns also emerged in deepfake behaviours: males were significantly more likely to have created or shared deepfake content.
Both genders improved in AI knowledge post-intervention, yet females showed a richer profile of gains: wider conceptual understanding, greater confidence, and meaningful increases in AI and CS career interest that partially narrowed the gender STEM gap. 
These findings highlight the need for gender-responsive AI curricula, particularly deepfake safety education for male students, and demonstrate that even single-day workshops can narrow gender gaps in STEM aspirations and AI confidence. 

\end{abstract}

\section{Introduction}

Integration of Artificial Intelligence (AI) into daily life has prompted calls for structured AI literacy education~\cite{Pedro2019ArtificialDevelopment,Schiff2021EducationStrategies}.
While emerging frameworks address what students should know about AI~\cite{Touretzky2019EnvisioningAI}, less attention has been paid to \emph{who} benefits from these interventions and which strategies are most effective for diverse demographic groups.
Gender disparities in computing education are well-documented~\cite{Kucuk2020StudentsExperience,Wang2017DiversitySocial}, yet empirical evidence on gender differences in AI literacy among secondary students remains sparse. This gap is consequential, as workshops become primary vehicles for AI literacy in schools~\cite{Renzella2025BuildingClassrooms}, we must understand whether they benefit all students equally to ensure equitable outcomes.
A recent systematic review found only one relevant study exploring the impact of gender on AI literacy,~\cite{Tan2025UnveilingResearch}, with research focusing mostly on tertiary populations~\cite{Cachero2025GenderStudy,Mansoor2024ArtificialSurvey}.
Moreover, the rapid rise of synthetic media (deepfake technology) creates gender-differentiated safety risks~\cite{Lazard2025DeepfakeReview}, yet the intersection of AI literacy education and deepfake awareness has received limited empirical attention.

This study evaluates \emph{\textbf{Anonymous program}}'s workshops, inspired by prior work which found overall improvements in AI knowledge but did not disaggregate results by gender~\cite{Bergh2026AIInterventions}. We address this gap with two research questions:

\smallskip
\noindent\textbf{RQ1.} What gender differences exist in baseline AI literacy, tool usage, safety awareness, and STEM career aspirations?

\noindent\textbf{RQ2.} Does the impact of an AI literacy workshop differ by gender?

\section{Method}

\subsection{Participants and Setting}

Participants were Years~7, 8 and~10 students from two regional, co-educational government secondary schools in New South Wales (NSW), Australia. These students completed pre- and post-intervention surveys as part of the \emph{\textbf{Anonymous program}}.
Single-sex schools and Year~9 were excluded to avoid a gender-school confound.
Paper surveys were administered anonymously and responses could not be paired.
Straight-line responses were removed ($n_{\text{pre}}=\nPreStraight$, $n_{\text{post}}=\nPostStraight$), resulting in $N_{\text{pre}} = \nPreTotal$ and $n_{\text{post}} = \nPostTotal$.
With a gender distribution: Male $n_{\text{pre}}=\nPreM$, $n_{\text{post}}=\nPostM$; Female $n_{\text{pre}}=\nPreF$, $n_{\text{post}}=\nPostF$. Students who identified as non-binary, preferred not to answer, or used a different gender label were excluded due to insufficient sample size ($n \leq 9$).

\subsection{Instrument}

The survey instrument was adapted from work described by Bergh et al.~\cite{Bergh2026AIInterventions}, covering nine constructs:
(1)~AI knowledge (5-point Likert: None--Expert);
(2)~understanding AI;
(3)~AI attitudes (5-point Likert: Disagree--Agree);
(4)~AI confidence (5-point Likert: Not at all--Extremely);
(5)~tool usage frequency (5-point Likert: Rarely/Never--Daily);
(6)~conceptual application; 
(7)~AI use motivations and knowledge sources;
(8)~deepfake\footnote{Defined as: "A video, images or audio that has been altered with AI to make it appear like someone said or did something that they did not".} experience (multi-select);
and
(9)~STEM career interest (5-point Likert: Not at all--Very interested).

\subsection{Analysis}

All analyses controlled for potential year level and school confounds.
Ordinary Least Squares (OLS) regression was used for all ordinal (Likert-scale) items following evidence that parametric methods are robust for Likert scales~\cite{SullivanAnalyzingScales}, with the model \texttt{outcome $\sim$ gender + year\_level + school}, with gender dummy-coded (Male = reference), and year level and school as categorical dummies.
Cohen's $d$ was computed as $\beta_{\text{gender}} / \sqrt{\text{MSE}}$ and interpreted using standard thresholds: $|d|<.20$ negligible, $<.50$ small, $<.80$ medium, $\geq.80$ large.
For binary items (multi-select questions), we used logistic regression with the same covariate structure, reporting Odds Ratios (OR) for the gender coefficient.
For RQ2, within-gender pre- and post-intervention comparisons used the model \texttt{outcome $\sim$ phase + year\_level + school}.
Uncorrected $p$-values are reported with effect sizes to highlight potential trends.

\section{Results}

\subsection{RQ1: Baseline Gender Differences}

\paragraph{Knowledge, understanding, and attitudes.}
Table~\ref{tab:rq1_ordinal} presents OLS regression results for ordinal items at baseline, controlling for year level and school.
Male students reported significantly higher self-assessed understanding of how AI systems are trained ($M_M=\rqiUndTrainMM$ vs.\ $M_F=\rqiUndTrainMF$, $d=\rqiUndTrainD$, $p=\rqiUndTrainP$).
The largest gender differences emerged in STEM career aspirations, with male students reporting significantly higher interest in engineering ($M_M=\rqiCareerEngMM$ vs.\ $M_F=\rqiCareerEngMF$, $d=\rqiCareerEngD$, $p=\rqiCareerEngP$), computer science ($M_M=\rqiCareerCSMM$ vs.\ $M_F=\rqiCareerCSMF$, $d=\rqiCareerCSD$, $p=\rqiCareerCSP$), and AI ($M_M=\rqiCareerAIMM$ vs.\ $M_F=\rqiCareerAIMF$, $d=\rqiCareerAID$, $p=\rqiCareerAIP$).
In the pre-workshop survey, no significant gender differences were found in AI knowledge, AI confidence, privacy awareness, data consent importance, or understanding AI bias after controlling for year level and school.

\begin{table}[b]
\caption{Baseline gender differences on Likert items via the OLS model.}
\label{tab:rq1_ordinal}
\centering
\small
\begin{tabular}{@{}lccccccc@{}}
\toprule
& \multicolumn{2}{c}{Male} & \multicolumn{2}{c}{Female} & & & \\
\cmidrule(lr){2-3} \cmidrule(lr){4-5}
Item & $M$ & $n$ & $M$ & $n$ & $\beta$ & $p$ & $d$ \\
\midrule
AI Knowledge           & \rqiAIKnowMM & \rqiAIKnownM  & \rqiAIKnowMF & \rqiAIKnownF & $\rqiAIKnowB$  & \rqiAIKnowP & $\rqiAIKnowD$ \\
AI Confidence          & \rqiAIConfMM & \rqiAIConfnM  & \rqiAIConfMF & \rqiAIConfnF & $\rqiAIConfB$  & \rqiAIConfP & $\rqiAIConfD$ \\
Understand AI Training    & \rqiUndTrainMM & \rqiUndTrainnM  & \rqiUndTrainMF & \rqiUndTrainnF & $\rqiUndTrainB$  & \rqiUndTrainP & $\rqiUndTrainD$ \\
Understand AI Bias/Errors        & \rqiUndBiasMM & \rqiUndBiasnM  & \rqiUndBiasMF & \rqiUndBiasnF & $\rqiUndBiasB$  & \rqiUndBiasP & $\rqiUndBiasD$ \\
%Privacy skills (Social Media) & \rqiPrivSocMM & \rqiPrivSocnM  & \rqiPrivSocMF & \rqiPrivSocnF & $\rqiPrivSocB$  & \rqiPrivSocP & $\rqiPrivSocD$ \\
%Privacy skills (Schoolwork)   & \rqiPrivSchoolMM & \rqiPrivSchoolnM  & \rqiPrivSchoolMF & \rqiPrivSchoolnF & $\rqiPrivSchoolB$  & \rqiPrivSchoolP & $\rqiPrivSchoolD$ \\
Data Consent Importance & \rqiConsentMM & \rqiConsentnM  & \rqiConsentMF & \rqiConsentnF & $\rqiConsentB$  & \rqiConsentP & $\rqiConsentD$ \\
Career: Artificial Intelligence             & \rqiCareerAIMM & \rqiCareerAInM  & \rqiCareerAIMF & \rqiCareerAInF & $\rqiCareerAIB$  & \rqiCareerAIP & $\rqiCareerAID$ \\
Career: Computer Science             & \rqiCareerCSMM & \rqiCareerCSnM  & \rqiCareerCSMF & \rqiCareerCSnF & $\rqiCareerCSB$  & \rqiCareerCSP & $\rqiCareerCSD$ \\
Career: Engineering    & \rqiCareerEngMM & \rqiCareerEngnM  & \rqiCareerEngMF & \rqiCareerEngnF & $\rqiCareerEngB$  & \rqiCareerEngP & $\rqiCareerEngD$ \\
\bottomrule
\end{tabular}
\end{table}

\paragraph{AI use motivations and tool usage.}
Females were significantly more likely to report using AI for schoolwork (\MotSchoolPctF\% vs.\ \MotSchoolPctM\%, OR=\MotSchoolOR, $p=\MotSchoolP$) and seeking advice (\MotAdvicePctF\% vs.\ \MotAdvicePctM\%, OR=\MotAdviceOR, $p=\MotAdviceP$).
The gender difference in problem-solving motivation was not significant after controlling for year level and school (\MotProblemPctF\% vs.\ \MotProblemPctM\%, OR=\MotProblemOR, $p=\MotProblemP$). No significant gender difference was observed in recreational AI use (\MotFunPctM\% vs.\ \MotFunPctF\%, OR=\MotFunOR, $p=\MotFunP$).
Regarding tool usage frequency, the largest gender difference was observed for Microsoft Copilot ($d=\CopilotD$, medium), used more frequently by males.

\paragraph{Deepfake experience and safety.}

Males and females reported having \textit{seen} deepfake content at similar rates (not capturing instances where students have unknowingly seen deepfake content) (Table~\ref{tab:deepfake}). While males were significantly more likely to have \textit{shared} a deepfake (\DfSharedPctM\% vs.\ \DfSharedPctF\%, OR=\DfSharedOR, $p=\DfSharedP$), females were significantly more likely to have \textit{reported} a deepfake (14.1\% vs.24.0\%, OR=2.14, $p=.046$). Additionally, males trended toward \textit{creating} a deepfake (\DfCreatedPctM\% vs.\ \DfCreatedPctF\%) or having been the \textit{subject} of one (\DfOfMePctM\% vs.\ \DfOfMePctF\%), though not statistically significant after controlling for year level and school.

\begin{table}[h]
\caption{Deepfake experience and AI use motivation by gender (pre-survey, logistic regression controlling for year level and school). Bold $p$ values: $p<.05$.}
\label{tab:deepfake}
\centering
\small
\begin{tabular}{@{}lcccl@{}}
\toprule
Item & Male (\%) & Female (\%) & OR & $p$ \\
\midrule
Seen deepfake content      & \DfSeenPctM & \DfSeenPctF & \DfSeenOR & \DfSeenP \\
Shared deepfake content    & \DfSharedPctM & \DfSharedPctF & \DfSharedOR & \textbf{\DfSharedP} \\
Created deepfake content   & \DfCreatedPctM &  \DfCreatedPctF & \DfCreatedOR & \DfCreatedP \\
Deepfake content of me     & \DfOfMePctM &  \DfOfMePctF & \DfOfMeOR & \DfOfMeP \\
%Friends creating deepfakes & \DfFriendsPctM &  \DfFriendsPctF & \DfFriendsOR & \DfFriendsP \\
%Know how to report         & \DfKnowReportPctM & \DfKnowReportPctF & \DfKnowReportOR & \DfKnowReportP \\
Have reported              & 14.1 & 24.0 & 2.14 & \textbf{.046} \\
%Know what to do if targeted & \DfKnowToDoPctM & \DfKnowToDoPctF & \DfKnowToDoOR & \DfKnowToDoP \\
\midrule
Use AI for schoolwork      & \MotSchoolPctM & \MotSchoolPctF & \MotSchoolOR & \textbf{\MotSchoolP} \\
Seek advice from AI        & \MotAdvicePctM & \MotAdvicePctF & \MotAdviceOR & \textbf{\MotAdviceP} \\
\bottomrule
\end{tabular}
\end{table}

\subsection{RQ2: Differential Intervention Effects}

Both genders showed statistically significant post-intervention gains in AI knowledge (Male: $d=\rqiiMAIKnowD$, medium; Female: $d=\rqiiFAIKnowD$, medium), and AI confidence (Male: $d=\rqiiMAIConfD$, medium), (Female: $d=\rqiiFAIConfD$, medium) (Table~\ref{tab:rq2}).
Female students gained across a wider range of items, including medium effects on understanding of AI training ($d=\rqiiFUndTrainD$) and AI bias ($d=\rqiiFUndBiasD$). Female students also showed a larger increase in the importance of consent when sharing data with AI ($d=\rqiiFConsentD$).
Most notably, female students showed significant medium-sized gains in career interest for both AI ($\Delta M=\rqiiFCareerAIDeltaM$, $d=\rqiiFCareerAID$) and computer science ($\Delta M=\rqiiFCareerCSDeltaM$, $d=\rqiiFCareerCSD$). Male students showed no significant change in career aspirations for either field.
As RQ1 identified baseline gender gaps in STEM career aspirations, these gains suggest the workshop may help narrow the gender--STEM gap.

\begin{table}[h]
\caption{Pre--post change by gender (OLS: outcome $\sim$ phase + year level + school). Bold $p$: significant at $\alpha=.05$. $d$ = Cohen's $d$.}
\label{tab:rq2}
\centering
\small
\begin{tabular}{@{}lccccccc@{}}
\toprule
& \multicolumn{3}{c}{Male ($n_{\text{pre}}$=\nPreM, $n_{\text{post}}$=\nPostM)} & \multicolumn{3}{c}{Female ($n_{\text{pre}}$=\nPreF, $n_{\text{post}}$=\nPostF)} \\
\cmidrule(lr){2-4} \cmidrule(lr){5-7}
Item & $\Delta M$ & $p$ & $d$ & $\Delta M$ & $p$ & $d$ \\
\midrule
AI Knowledge      & \rqiiMAIKnowDeltaM & \rqiiMAIKnowP & \rqiiMAIKnowD & \rqiiFAIKnowDeltaM & \rqiiFAIKnowP & \rqiiFAIKnowD \\
AI Confidence     & \rqiiMAIConfDeltaM & \rqiiMAIConfP & \rqiiMAIConfD & \rqiiFAIConfDeltaM & \rqiiFAIConfP & \rqiiFAIConfD \\
Understand AI Train. & \rqiiMUndTrainDeltaM & \rqiiMUndTrainP & \rqiiMUndTrainD & \rqiiFUndTrainDeltaM & \rqiiFUndTrainP & \rqiiFUndTrainD \\
Understand AI Bias   & \rqiiMUndBiasDeltaM & \rqiiMUndBiasP & \rqiiMUndBiasD & \rqiiFUndBiasDeltaM & \rqiiFUndBiasP & \rqiiFUndBiasD \\
%Privacy (Social Media)  & \rqiiMPrivSocDeltaM & \rqiiMPrivSocP & \rqiiMPrivSocD & \rqiiFPrivSocDeltaM & \rqiiFPrivSocP & \rqiiFPrivSocD \\
%Privacy (Schoolwork) & +0.22 & .125 & .252 & +0.24 & .105 & .257 \\
Data Consent Import.   & \rqiiMConsentDeltaM & \rqiiMConsentP & \rqiiMConsentD & \rqiiFConsentDeltaM & \rqiiFConsentP & \rqiiFConsentD \\
Career: AI        & \rqiiMCareerAIDeltaM & \rqiiMCareerAIP & \rqiiMCareerAID & \rqiiFCareerAIDeltaM & \rqiiFCareerAIP & \rqiiFCareerAID \\
Career: CS        & \rqiiMCareerCSDeltaM & \rqiiMCareerCSP & \rqiiMCareerCSD & \rqiiFCareerCSDeltaM & \rqiiFCareerCSP & \rqiiFCareerCSD \\
Career: Eng.      & \rqiiMCareerEngDeltaM & \rqiiMCareerEngP & \rqiiMCareerEngD & \rqiiFCareerEngDeltaM & \rqiiFCareerEngP & \rqiiFCareerEngD \\
\bottomrule
\end{tabular}
\end{table}

Both genders showed an increase in recognition of AI's role within entertainment platforms when identifying platforms that use AI, specifically Netflix (Male: $\rqiiIDMNetflixChange\%$, $p=\rqiiIDMNetflixP$; Female: $\rqiiIDFNetflixChange\%$, $p=\rqiiIDFNetflixP$) and Spotify (Male: $\rqiiIDMSpotifyChange\%$, $p=\rqiiIDMSpotifyP$; Female: $\rqiiIDFSpotifyChange\%$, $p=\rqiiIDFSpotifyP$).
Females also showed gains in identifying TikTok as a platform that uses AI ($\rqiiIDFTikTokChange\%$, $p=\rqiiIDFTikTokP$).

\section{Discussion and Conclusion}

We provide, to our knowledge, the first gender-disaggregated analysis of an AI literacy workshop with Australian secondary students. Key findings emerge:

First, AI literacy workshops may help narrow the gender--STEM career gap. At baseline, male students reported higher interest in AI, computer science, and engineering careers. Post-intervention, female students showed significant gains in career interest in AI and computer science, while male interest remained static. Despite these gains, neither group's average response reached the "interested" level on the scale across all three career domains.

Second, the workshop produced gender-differentiated outcomes. Both genders gained in self-reported AI knowledge and confidence, but females showed broader gains, notably in conceptual understanding of AI training, and biases in AI.

Third, deepfake behaviours show gender patterns. Males were more likely to have shared deepfake content (\DfSharedPctM\% vs.\ \DfSharedPctF\%), and showed trends toward higher rates of involvement, though the survey did not differentiate between benign or harmful deepfake content (e.g. sexualised imagery). Interestingly, female students were more likely to have reported deepfakes (14.1\% vs 24.0\%). Findings suggest that targeted safety intervention for male students regarding the ethical dissemination of synthetic media is necessary and defensible.

\paragraph{Limitations.}
The sample is restricted to two co-educational schools in regional NSW, limiting generalisability. The anonymous survey design prevents paired analyses and may be subject to self-reporting bias. Future research should investigate types of deepfakes, creation of, sharing, and victimisation of, by gender.

\paragraph{Implications.}
We present three recommendations for AI literacy program design: (1)~leverage  workshops to increase female STEM career interest; (2)~incorporate gender-specific deepfake safety modules, particularly for male students; and (3)~ensure that workshop evaluations monitor equity of impact by gender.

\begin{credits}
\subsubsection{\ackname} Day of AI Australia and its research activities are funded by philanthropic donations and corporate partnerships.

\subsubsection{\discintname} 
Institution 2 is a registered Australian charity. Institution 1 and 2 have a legal collaboration. Institution 2 has previously received funding from Google.org and Microsoft, contributions have not funded this research.

%\begin{credits}
%\subsubsection{\ackname} Space reserved for acknowledgements, space reserved for acknowledgements space reserved for acknowledgements.

%\subsubsection{\discintname} 
%Space reserved for disclosure, space reserved for disclosure, space reserved for disclosure space reserved for disclosure space reserved for disclosure.
\end{credits}

\bibliographystyle{splncs04}
\bibliography{references}
\end{document}